\documentstyle [twocolumn,amssymb,epsf]{mn}
\newcommand{\mincir}{\raise
-3.truept\hbox{\rlap{\hbox{$\sim$}}\raise4.truept\hbox{$<$}\ }}
\newcommand{\magcir}{\raise
-3.truept\hbox{\rlap{\hbox{$\sim$}}\raise4.truept\hbox{$>$}\ }}
\newcommand{\minmag}{\raise
-3.truept\hbox{\rlap{\hbox{$<$}}\raise5.truept\hbox{$<$}\ }}
\newcommand{\be}{\begin{equation}}
\newcommand{\ee}{\end{equation}}
\newcommand{\ba}{\begin{eqnarray}}
\newcommand{\ea}{\end{eqnarray}}
\newcommand{\brr}{\begin{array}}
\newcommand{\err}{\end{array}}
\newcommand{\bc}{\begin{center}}
\newcommand{\ec}{\end{center}}

\renewcommand{\baselinestretch}{1.0}

\title[Comparison between optical and X-ray cluster detection
methods]{The XMM-{\em Netwon}/2dF survey III:
Comparison between optical and X-ray cluster detection
methods}
\author[Basilakos et al.]
{S. Basilakos$^{1}$, M. Plionis$^{1,2}$, A. Georgakakis$^{1}$, 
I. Georgantopoulos$^{1}$, T. Gaga$^{1,3}$, 
\newauthor
V. Kolokotronis$^{1}$, G. C. Stewart$^{4}$ 
\\
\vspace{0.1cm}
$^1$ Institute of Astronomy \& Astrophysics, National Observatory of Athens, 
I.Metaxa \& B.Pavlou, Palaia Penteli, 152 36, Athens, Greece \\
$^{2}$ Instituto Nacional de Astrofisica, Optica y Electr\'onica (INAOE)
Apartado Postal 51 y 216, 72000, Puebla, Pue., Mexico \\
$^3$ Physics Department, Univ. of Athens, Panepistimioupolis,
Zografou, Athens, Greece \\
$^4$ Department of Physics and Astronomy, University of Leicester, 
UK, LE1 7RH\\
}

\begin{document}

\maketitle

\begin{abstract}
We directly compare X-ray and optical techniques of cluster detection
by combining SDSS photometric data with a wide-field ($\sim 1.6$
deg$^{2}$) XMM-{\em Newton} survey near the North Galactic Pole region.
The optical cluster detection procedure is based on 
merging two independent selection methods - a smoothing+percolation    
technique, and a Matched Filter 
Algorithm. The X-ray cluster detection is based on a wavelet-based algorithm,
incorporated in the SAS v.5.3 package. The final optical sample 
counts nine candidate clusters with estimated APM-like richness of
more than 20 galaxies, 
while the X-ray based cluster candidates are four. 
Three out of these four X-ray cluster candidates are also optically detected.
We argue that the cause is that the majority of the optically detected
clusters are relatively poor X-ray emitters, with X-ray fluxes 
fainter than the flux limit (for extended sources) of our survey 
$f_{x}(0.3-2 {\rm keV}) \simeq 2 \times 10^{-14} {\rm erg ~cm^{-2}~s^{-1}}$.

{\bf Keywords:} galaxies: clusters: general - cosmology:observations 
- cosmology: large-scale structure of Universe.
\end{abstract}

\section{Introduction}

Clusters of galaxies occupy an eminent position in the 
structure hierarchy, being the most massive virialized systems 
known and therefore they appear to be ideal tools 
for testing theories of structure formation and extracting 
cosmological informations (cf. Bahcall 1988; West Jones \& Forman 1995; 
B\"{o}hringer 1995; Carlberg et al. 1996; Borgani \& Guzzo 2001;  
Nichol 2002 and references therein). 

To investigate the global properties of the 
cosmological background it is necessary to construct and study
large samples of clusters (cf. Borgani \& Guzzo 2001). This
understanding has initiated a number of studies aiming to compile
unbiased cluster samples to high redshifts, utilizing multiwavelength data
(e.g. optical, X-ray, radio). 
On the other hand, the study of individual clusters, 
provide complementary information regarding their physical properties
and evolutionary processes.
Overall, it is very important to fully
understand the selection effects that enter in 
the construction of cluster samples since these could bias
any statistical analysis of these samples (cf. Sutherland 1988).

At optical wavelengths there are several available 
samples in the literature. For example, the Abell/ACO catalogue   
(Abell, Corwin \& Olowin 1989) was constructed by visual inspection 
of the Palomar Observatory Sky Survey plates 
and is still playing an important role in 
astronomical research. Since then, a large number of 
optically selected samples constructed with automated methods 
have been constructed: EDCC ({\em Edinburgh Durham Cluster Catalogue}; 
Lumsden et al. 1992), APM ({\em Automatic Plate Measuring}; 
Dalton et al. 1994), PSCS ({\em Palomar Distant Cluster Survey}; Postman et al. 1996),
EIS ({\em ESO Imaging Cluster Survey}; Olsen et al. 1999),
RCS {\em Red-Sequence Cluster Survey}; Gladders \& Yee 2000) 
and the Sloan Digital Sky cluster survey (Goto et al. 2002; Bahcall et al. 2003). The above
cluster samples, based on different selection methods, 
aim to obtain homogeneously selected optical cluster samples with 
redshifts that extended beyond the $z\sim 0.2$ limit of the Abell/ACO 
catalogue. We should mention that 
the advantage of using optical data is the shear size of 
the available cluster catalogues and thus the 
statistical significance of the emanating results.

A major problem here is that the optical surveys suffer from severe systematic
biases which are due to projection effects. Background 
and foreground galaxies, projected  
on the cluster could distort the identifications (e.g. Frenk et al. 1990),
which is particularly true for poor systems at high redshifts. X-ray 
surveys provide an alternative method for compiling cluster
samples, owing to the fact that the diffuse
Intra-Cluster Medium (ICM) emits strongly at X-ray wavelengths. This emission
is proportional to the square of the hot gas density,
resulting in a high contrast with respect to the unresolved X-ray
background, and thus X-ray selected clusters 
are less susceptible to projection effects.
Therefore the main advantage, inherent in the X-ray
selection of flux-limited samples 
is that the survey volume
and hence number densities, luminosity and mass
functions can be reliably computed. Furthermore,
X-ray cluster surveys can be used to study
cluster dynamics and morphologies, the Sunyaev 
Zeldovich effect and finally their cosmological evolution.

The first such sample, with large impact to the studies of clusters,
was based on the Extended Einstein Medium 
Sensitivity Survey, containing 99 clusters (Stocke et al. 1991). 
The {\it ROSAT} satellite with its large field of view (FOV) 
and better sensitivity, allowed
a leap forward in the X-ray cluster astronomy, producing large samples
of both nearby and distant clusters (e.g. Castander et al. 1995; Ebeling 
et al. 1996a, 1996b; Scharf et al. 1997; Ebelling et al. 2000; 
B\"{o}hringer et al. 2001; Gioia et al. 2001; 
B\"{o}hringer et al. 2002; Rosati, Borgani \& Norman 2002
and references therein). Recently, the XMM-{\it Newton} observatory 
with $\sim 10$ times more effective area and $\sim 5$ times better spatial 
resolution than the {\it ROSAT} provides an ideal platform for the
study of galaxy clusters.

However, even with the improved sensitivity of the XMM-{\it Newton}, optical
surveys still remain significantly more efficient and less expensive in 
telescope time for compiling cluster samples, albeit with the
previously discussed limitations (eg. incompleteness, projection
effects etc). Therefore it is necessary to study the different
selection effects and biases that enter in detecting clusters in the
two wavelength regimes.
Donahue et al. (2002) using the {\it ROSAT} Optical
X-ray Survey (ROXS), found that using both X-ray and optical 
methods to identify clusters of galaxies, the overlap was poor. About
$20\%$ of the optically detected clusters were found in 
X-rays while 60\% of the X-ray clusters were identified also in the optical
sample. Furthermore, not all of their X-ray detected clusters
had a prominent red-sequence, a fact that could
introduce a bias in constructing cluster samples using only
colour information (as in Goto et al. 2002).

The aim of this work is along the same lines, attempting 
to make a detailed comparison of optical and X-ray cluster
identification methods in order to quantify the selection biases
introduced by these different techniques.
We use XMM-{\em Newton} which has a factor of $\sim 5$ better spatial
resolution and an order of magnitude more effective area at 1 keV,
making it an ideal instrument for the detection of relatively distant
clusters.

The plan of the paper is as follows. The observed data sets are 
presented in Section 2. In Section 3 we discuss  the 
methods employed to identify candidate optical clusters and comment on the
systematic effects introduced in our analysis. 
Also, Section 3 describes our projected cluster shape determination method
as well as the cluster surface brightness based on a King like profiles. 
In Section 4 we compare the optical and X-ray selected 
cluster samples. Finally, in Section 5, we present our conclusions.
Throughout this paper we use $H_{\circ}=100h$ km s$^{-1}$Mpc$^{-1}$ 
and $\Omega_{\rm m}=1-\Omega_{\Lambda}=0.3$.

\section{Observations}

\subsection{The optical data}
In our analysis we use the SDSS Early Data Release (EDR), covering an
area of $\sim 400$deg$^2$ in the sky (Stoughton et al. 2002). The SDSS 
is an ongoing imaging and spectroscopic survey that  
covers $\sim$10000 deg$^{2}$ of the sky. Photometry is 
obtained in five bands ($u, g, r, i, z$) to the limiting 
magnituge $r \lesssim 22.5 \;{\rm mag}$,
providing an homogeneous multi-color photometric catalogue.

Goto et al. (2002), applied an object cluster finding algorithm to the
SDSS EDR and produced a list of galaxy clusters (hereafter CE catalogue),
with estimated photometric redshifts which contains 2770 and 
1868 galaxy clusters in the North and South slices, respectively.
The cluster redshifts were estimated
using color information by identifying the bin in $g - r$ which 
has the largest number of galaxies around the color prediction of
elliptical galaxies (Fukugita, Shimasaku, Ichikawa 1995)
at different redshifts (which define the different
$g - r$ bins). Note that the true and estimated
redshifts are better correlated for $z<0.3$, 
with rms scatter $\sim 0.015$, while for
$z>0.3$ it is $\sim 0.021$. Note, that the CE method is optimized up
to $z \leq 0.4$ and becomes highly insensitive at higher redshifts.


\subsection{The XMM-{\em Newton} data}
We analyzed 9 {\it XMM-Newton} fields with nominal exposure times
between 2 and 10 ksec. 
The {\it XMM-Newton} field-of-view is a circle with a radius
of 15 arcmin and thus our survey covers an area of $\rm 1.8 \;\;deg^2$. 
However, one of the fields, suffering from significantly elevated and 
flaring particle background, was excluded from the 
X-ray analysis. This reduces our effective area to $\rm 1.6 \;\;deg^2$. 
 The details of the X-ray observations are given
in Georgakakis et al. (2004) and Georgantopoulos et al. {\em 
in preparation}. 

\begin{figure}
\mbox{\epsfxsize=8cm \epsffile{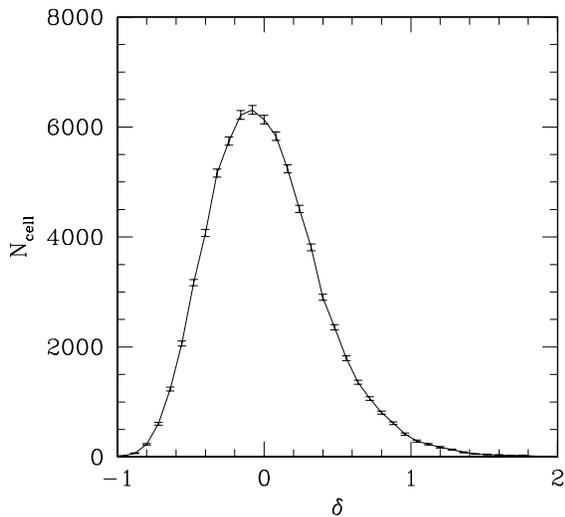}}
\caption{The SDSS probability density function ({\em pdf}).}
\end{figure}

\section{Optical Cluster Detection}
Identifying real clusters using imaging data is a difficult
task since projection effects can significantly affect the visual
appearance of clusters. Many different
algorithms have been proposed and applied on different data sets. In
what follows we present a new cluster finding algorithm (based on
two distinct algorithms) and apply it
on the subsample of the $r$-band 
SDSS data\footnote{We have tested that our results remain
qualitatively unaltered when using the $i$-band.}, that overlaps
with our XMM-{\em Netwon} survey.

\subsection{The Smoothing+Percolation Procedure}
This cluster detection algorithm (hereafter SMP method) 
is based on smoothing the discrete distribution using
a Gaussian smoothing kernel on a $N_{gr}\times N_{gr}$ grid:
\begin{equation}\label{eq:ker}
{\cal W}(|{\bf x}_{i}-{\bf x_{gr}}|)  = \frac{1}{\sqrt{2 \pi R_{\rm sm}^{2}} }
\exp\left(-\frac{|{\bf x}_{i}-{\bf x_{gr}}|^{2}}{2 R_{\rm sm}^{2}}
\right) \;,
\end{equation}
with $R_{\rm sm}$ the smoothing radius in grid-cell units and ${\bf
  x}_i$ the Cartesian position of the $i^{th}$ galaxy.
The smoothed surface density, of the $j^{th}$ cell at the grid-cell positions
$\bf{x_ {gr}}$, is:
\begin{equation}\label{eq:rhog}
\rho_{j}({\bf x_{gr}}) = \frac{\sum_{i} \rho_{j}({\bf x}_{i})
{\cal W}(|{\bf x}_{i}-{\bf x_{gr}}|)}
{\int {\cal W}(|{\bf x_{i}}-{\bf x_{gr}}|) {\rm d}^{2}x},
\end{equation}
where the sum is over the distribution of galaxies with positions
${\bf x}_{i}$. For each cell we define its density fluctuation,
$\delta_j$, as:
\begin{equation}
\delta_j = \frac{\rho_{j}(x_{gr})-\langle \rho \rangle }{\langle \rho
\rangle} \;,
\end{equation}
where $\langle \rho \rangle$ is the mean projected SDSS galaxy density
and the probability density function, $f(\delta)$, which is plotted in Fig.~1.
We select all grid-cells with overdensities 
above a chosen critical threshold  ($\delta \ge \delta_{\rm cr}$)
and then we use a friends-of-friends algorithm to link all adjacent
cells in order to form groups of connected cells, which we consider as
our candidate clusters.
Note that the grid cell size is such that at $z=0.4$ it
corresponds to 100 $h^{-1}$kpc ($\sim 19''$).

There are three free parameters in the above procedure: 
the critical overdensity threshold ($\delta_{\rm cr}$)
the smoothing radius ($R_{\rm sm}$) and the
number of connected cells ($n_c$) above which we consider a candidate cluster.
Due to the fact that clusters should be identified 
as high density regions, we choose the critical value of the
overdensity to be $\delta_{\rm cr}\simeq 1$ which corresponds to the
$\sim 98$ percentile
of the probability density function ({\em pdf}) 
plotted in Fig.1. Had we used lower values of 
$\delta_{\rm cr}$ we would have percolated overdense regions through
large parts of the whole area.
\begin{figure}
\mbox{\epsfxsize=7.5cm \epsffile{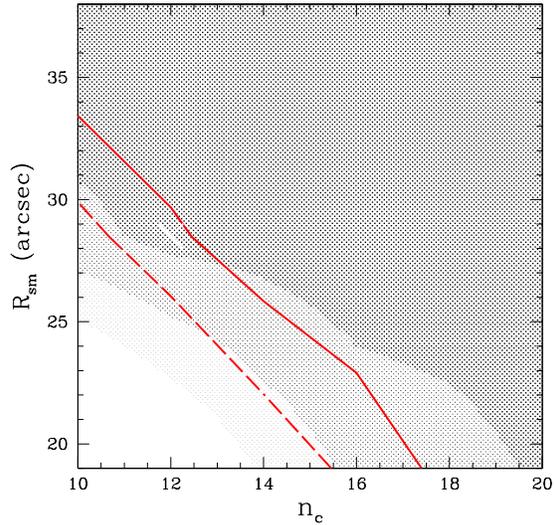}}
\caption{Statistical significance of our cluster detection procedure as a
function of the smoothing radius $R_{\rm sm}$ and minimum number of
connected cells, $n_c$. The darkest greyscale corresponds to ${\cal
  P}=0.96$, while the faintest to ${\cal P}<$0.88 . The continuous
line corresponds to the expected number of APM clusters, while the
broken line to the number Goto et al. (2002) clusters. The cross
reflects our best choice of the free parameters ($R_{\rm sm}$ and $n_c$).}
\end{figure} 

The $R_{\rm sm}$ and $n_c$ parameters were chosen in such a way as
to minimize the number of spuriously detected clusters. To this end we
perform a series of Monte-Carlo simulations in which we randomize the
positions of the SDSS galaxies, destroying its intrinsic clustering,
while keeping unchanged the galaxy redshift selection function.
On this intrinsically random galaxy distribution we apply
the procedure described above, by varying
the values of the free parameters ($R_{\rm sm}$ and $n_c$).
We define the probability of detecting real clusters 
in the SDSS data, as
${\cal P}(R_{\rm sm}, n_c) = 1-N_{\rm rand}/N_{\rm SDSS}$, 
were $N_{\rm ran}$ is the number of clusters detected in the
randomized distribution and $N_{\rm SDSS}$ is the number of clusters
detected in the true SDSS data.
In Fig. 2 we show as a greyscale these probabilities. Although,
increasing $R_{\rm sm}$ and $n_c$, results in very high probabilities
of detecting real high-density regions, it also results in 
small numbers of detected clusters. We therefore attempt to break this degeneracy by
using the expected number of clusters from existing
surveys. For example, the continuous line in Fig.2 corresponds to the
expected number of APM clusters in the area covered by our XMM/2dF
survey\footnote{This number ($\sim 40$) is estimated, without
assuming any evolution, by multiplying the 
local ($z<0.1$) mean number density of APM
clusters with the volume element covered by our survey out to a $z\sim
0.45$ within the
concordance cosmological model ($\Omega_\Lambda=0.7$).}, 
while the thick broken line delineates the corresponding
number of the Goto et al. (2002) clusters. 
The region of the highest probability, that falls within the above
cluster number limits, is that corresponding to:
$$R_{\rm sm} = 28.5^{''} \;\;\;\;\;\;\; n_{c} =12
\;,$$ 
which we choose as our optimal parameters for the detection of real
clusters. 
\begin{figure}
\mbox{\epsfxsize=7.5cm \epsffile{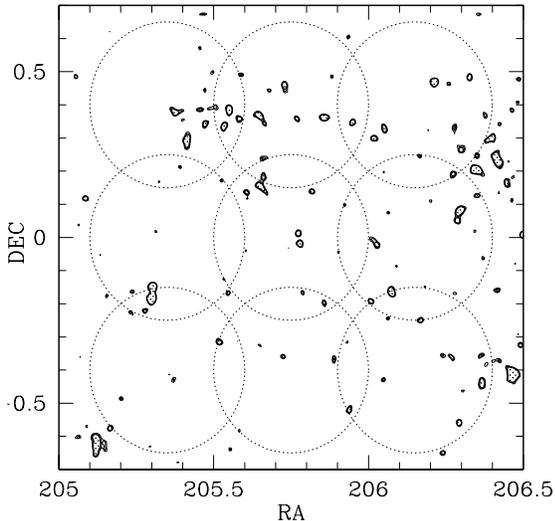}}
\caption{
The smooth SDSS density field on equatorial coordinates. 
The $\delta_{\rm cr}=1$ level appears as a thick continuous 
line. The large dotted circles represents the XMM 15 arcmin radius fields of
view of our shallow XMM/2dF survey.}
\end{figure}

In Fig. 3 we plot the smoothed SDSS galaxy distribution, using the
above value of $R_{\rm sm}$, on the equatorial plane with a contour 
step of $\Delta\delta=0.2$, starting from $\delta_{\rm cr} =1$.
Many overdense regions are apparent. Finally, we note that 
our cluster finding algorithm 
with $n_c=12$ produces a total of 25 candidate clusters, 
11 of which are common with the Goto et al. (2002) sample of 29
clusters in the same region.


\subsection{The Matched Filter Algorithm}
In addition to the SMP technique we have also employed 
the matched filter algorithm (hereafter MFA) described by 
Postman et al. (1996) to identify optical galaxy overdensities. 
The advantage of this method is that it exploits both positional and
photometric information producing galaxy density maps where
spurious galaxy fluctuations are suppressed. Also, 
an attractive feature of this method is that it provides redshift
estimates for the detected candidate galaxy clusters.
A drawback of the matched filter method is that one must assume a form
for the cluster luminosity function and radial profile. Clusters with
the same richness but different intrinsic shape or different
luminosity function from the assumed ones do not have the same
likelihood of being  detected.

A detailed description of the matched filter algorithm can be found in
Postman et al. (1996). In brief, the filter used to convolve the
galaxy catalogue is derived from an approximate maximum likelihood
estimator obtained from a model of the spatial and luminosity
distribution of galaxies within a cluster. At any given patch of the
sky the number density of galaxies per magnitude bin is

\begin{equation}\label{eq_1}
D(r,m)=b(m)+\Lambda_{cl}\,P(r/r_{c})\,\phi(m-m^{*}) \;\;,
\end{equation}
where $D(r,m)$ is the galaxy surface density
(cluster+field galaxies) at a given magnitude $m$ and at a distance
$r$ from the galaxy  cluster center, $b(m)$ is the field galaxy 
surface density, $P(r/r_{c})$ is the cluster projected radial
profile, $\phi(m-m^{*})$ is the cluster luminosity function and
$\Lambda_{cl}$ is an estimator of the cluster richness. The parameters
$r_{c}$ and $m^{*}$ are the characteristic cluster scale length and
apparent magnitude corresponding to the characteristic luminosity of
the cluster luminosity function. Postman  et al. (1996) showed that
the filter for cluster detection should maximize the expression

\begin{equation}\label{eq_2}
\int\,P(r/r_{c})\,\frac{\phi(m-m^{*})}{b(m)}\,D(m,r)\,d^{2}r\,dm \;\;.
\end{equation}
Assuming that the galaxy distribution $D(m,r)$ can be
represented by a series of $\delta$-functions at the observed
positions and magnitudes, the integral above is reduced to

\begin{equation}\label{eq_3}
S(i,j)=\sum_{k=1}^{N_T}\,P(r_k/r_c)\,L(m_k) \;\;,
\end{equation}
where $L(m_{k})$ is defined as

\begin{equation}\label{eq_4}
L(m)=\frac{\phi(m-m^{*})}{b(m)} \;\;.
\end{equation}
The quantity $S(i,j)$ is the value of the $(i,j)$ pixel of
the convolved galaxy density map and $N_T$ is total number of galaxies in
the catalogue.  $L(m_k)$ is the luminosity
weighting function (i.e. flux filter). In practice, the survey  area is
binned into pixels $(i,j)$ of given size (for the choice of values see
below) and the sum in equation \ref{eq_3} is evaluated by iterating
through all galaxies in the catalogue. This then  is repeated for
every pixel of the density map. Both $m^{*}$ and
$r_{c}$ are a function of redshift and hence $S(i,j)$ also depends on
redshift through these parameters. The redshift dependence of $m^{*}$
also includes a $k$-correction as discussed in the next section. 

The flux filter in equation \ref{eq_4} has a divergent integral at
faint magnitudes for Schechter luminosity functions with
$\alpha<-1$. To overcome this problem Postman et al. (1996) modified
the flux filter by introducing a power-law cutoff of the form
$10^{-0.4\,(m-m^*)}$. Therefore, the flux filter in equation
\ref{eq_3} is defined as:
\begin{equation}\label{eq_5}
L(m)=\frac{\phi(m-m^{*})\,10^{-0.4\,(m-m^*)}}{b(m)} \;\;.
\end{equation}
The radial filter in equation \ref{eq_3} is defined:
\begin{eqnarray}\label{eq_6}
P(r)= &\frac{1}{\sqrt{1+(r/r_c)^2}}-
\frac{1}{\sqrt{1+(r_{co}/r_{c})^2}}, &\mathrm{if \,\,\,r<r_{co}} \nonumber \\
& 0 & \mathrm{otherwise,}
\end{eqnarray}
where $r_c$ is the cluster core radius and $r_{co}$ is an
arbitrary cutoff radius. Here, we assume $r_{c}=100\,h^{-1}$\,kpc and
$r_{co}=1\,h^{-1}$\,Mpc. The characteristic absolute magnitude of the
luminosity function, in $r-$band is taken 
to be $M^{*}=-20.83+5{\rm log}h$ (Blanton
et al. 2003). Both the radial and the flux filter are normalized as 
described in Postman et al. (1996). 

\subsubsection{The MFA Cluster detection}
The matched filter algorithm is applied to a $\rm 1.6$ deg$^{2}$ subregion
of the SDSS centered on our XMM/2dF survey. We consider galaxies with
magnitudes brighter than $r= 22.5$\,mag. At fainter magnitudes  
the SDSS is affected by incompleteness. 
The galaxy density map $S(i,j)$ representing the galaxy density map is
independently estimated for redshifts between $z_{min}=0.1$ to
$z_{max}=0.6$, incremented in steps of 0.1. The $z_{max}$ corresponds to
the redshift where  $m^{*}$ becomes comparable to the limiting
magnitude of the survey.  The characteristic luminosity, $L^{*}$, the 
faint end slope of the luminosity function, $\alpha$, and the cluster
core radius, $r_{c}$,  are assumed to remain constant with
redshift. 

The conversion from luminosity to apparent magnitude requires an assumption 
to be made on the $k$-correction of the galaxies. Here we assume a
non-evolving elliptical galaxy model obtained from the
Bruzual \& Charlot (1993) stellar population synthesis code (Pozzeti,
Bruzual \& Zamorani 1996). It is important to(hereafter SMP method) 
emphasize that the choice of $k$-correction does not significantly affect
the cluster detections, but has an impact on the redshift determination. 

The pixel size of the galaxy density maps at any redshift is taken to be
$\approx19^{\prime\prime}$ corresponding to a projected cluster core
radius of $r_{c}=100\,h^{-1}$\,kpc at the redshift $z=0.4$ (see section 3.1). 
Galaxy density maps are created for redshifts 0.1--0.6 and stored in FITS
images. The peaks of the galaxy distribution are detected using
SExtractor (Bertin \& Arnouts 1996). The mean and the variance
of the background are determined using a global value in each
likelihood map. The main input parameters are the detection threshold,
$\sigma_{\rm det}$, given as a multiple of background variance, and the
minimum number of pixels, $N_{\rm min}$ for a peak  to be extracted as
candidate cluster. Simulations have been carried out  (see next
section) to optimize these parameters and to minimize the number 
of spurious cluster detections ($\approx5\%$). We adopt
$\sigma_{\rm det}=4.0$ while $N_{\rm min}$ is set to the area of a circle
with radius $r_{c}=100\,h^{-1}$\,kpc at any  given redshift.

The significance of a detection, $\sigma$, is defined as the maximum
value of all the pixels associated with that detection (i.e. pixels with
values above the SExtractor detection threshold), expressed in multiples of 
the rms noise above the background. The peaks (i.e. candidate clusters) 
extracted on density maps  of different redshifts are matched by positional
coincidence using a radius of $\approx1$\,arcmin.  The approximate
redshift of the candidate cluster is taken to be the redshift where
the detection significance is maximum. 

Within the  area of our XMM/2dF
survey the MFA method identifies 12 cluster candidates out of which 6
are found also by Goto et al. (2002) and 9 are found by the SMP method.

\subsubsection{The MFA simulations}
Monte Carlo simulations similar to those described in section 3.2 were
employed to test the performance of the matched filter algorithm and
to establish the best choice of extraction parameters. A total of 100
random galaxy catalogues over an area of $\sim 1.6$ deg$^{2}$
(i.e. similar to that covered by the real catalogue) were produced with
the same magnitude distribution as the real data-set. The matched
filter algorithm was then applied to these mock catalogues and  the
SExtractor was used to identify peaks in the  derived density maps.  

Following Olsen et al. (1999) we minimize the number of spurious
detections using the SExtractor parameters  $N_{\rm min}$ and
$\sigma_{\rm det}$ as well as other properties of the noise peaks
such as the number of redshifts where a peak is detected, $n_{z}$.
We adopt $\sigma_{\rm det}=4$, $N_{\rm min}=\pi\,r_{c}^{2}$ at any
given redshift and we only consider  peaks that appear in
at least two consecutive redshifts (i.e. $n_z\geq 2$). For this choice
of parameters we estimate the spurious cluster contamination
to be $\approx5\%$.  

\begin{table*}
\caption[]{List of the SDSS clusters in the XMM area. The correspondence 
of the columns is as follows:
index number, right ascension $\alpha$ and declination $\delta$ of the 
cluster center, cluster ellipticity $\epsilon$ and $\phi$ is the cluster position angle (in degrees),
the cross correlation results (C-C)- the numbers correspond to: 1 - CE method, Goto et al. (2002); 2 - 
Matched Filter Algorithm; 3 - X-ray extended sources (Gaga et al. {\em
  in preparation}); 4 - Couch et al. (1991). 
The next columns are: the estimated redshift by different methods, 
the core radius with its $2\sigma$ error and
the $\chi^{2}$ probabilities ($P_{\chi^{2}}$) of consistency between 
the intrinsic cluster density profile and King 
model. Finally, ${\cal N}_{g}$, $m_{3}$ and $f_{x}$ are the number of
galaxies, the magnitude of the third-brightest cluster member (within
a radius $\sim 0.75 \;h^{-1}$ Mpc)
and the detected flux or the $3\sigma$ upper flux limits in units of 
$\rm 10^{-14} erg \; cm^{-2} \;s^{-1}$ respectively (for details see
section 4).
Note that the $r_{c}$ has units of $h^{-1}$Mpc and the King profile 
slope used $\alpha \simeq 0.7$.}

\tabcolsep 7pt
\begin{tabular}{cccccccccccc} 
\hline
Index & $\alpha$ & $\delta$ & $\epsilon$ & $\phi$ & 
C-C & $z$ & $r_{c}$ &$P_{\chi^{2}}$ & ${\cal N}_{g}$ & $m_3$ & $f_{x}$ \\ \hline  
 1 & 205.513 &-0.316 & 0.10 & 113.7 &  2 &$0.10^{2}$   & $0.10\pm
 0.03$&0.14& 29& 17.38 & $<$0.40\\
 2 & 205.417 & 0.287 & 0.56 &   6.3 & 1,2,3   &$0.39^{1}-0.40^{2}$   &
 $0.50\pm 0.17$&0.003 & 26 & 18.96 & 1.86\\
 3 & 205.307 &-0.183 & 0.61 &   4.7 & 1,2 &  $0.40^{1}-0.60^{2}$ &
 $0.39 \pm 0.13$&0.73 &34& 19.37 & $<$0.16\\
 4 & 205.550 & 0.378 & 0.37 & 170.4 & 1,2& $0.31^{1}-0.50^{2}$ &
 $0.52\pm 0.17$ &0.95 &49& 19.76 & $<$3.91\\
 5 & 205.859 & 0.354 & 0.37 &  90.7 & 2&  $0.60^{2}$ &  $0.56 \pm
 0.19$&0.50 &50& 20.06 & $<$6.37\\
 6 & 205.777 &-0.022 & 0.10 & 158.2 & 2,3,4 &  $0.60^{2}-0.67^{4}$ &
 $0.35\pm 0.11$ &0.98&33& 20.03 & 5.55\\
 7 & 206.296 & 0.078 & 0.56 &  22.0 & 1,2&$0.37^{1}-0.30^{2}$ &
 $0.42\pm 0.14$ &0.59 &30 & 19.25 & $<$7.29\\
 8 & 206.363 &-0.438 & 0.43 &   3.3 & 2&   $0.30^{2}$&  $0.46 \pm
 0.15$&0.63 &29& 19.49 &$<$3.29\\
 9 & 206.210 & 0.468 & 0.10 &  48.0 & 1,2,3&   $0.36^{1}-0.30^{2}$&
 $0.09 \pm 0.03$&0.46 &28& 19.56 & 6.03\\
\hline
\end{tabular}
\end{table*}

\subsection{Our final optically selected cluster sample}
We construct our final optical cluster catalogue by adopting the conservative
approach of considering as cluster candidates those identified by
both the SMP and MFA techniques. Our final sample, which we call SMPMFA, 
consists of 9 clusters with SDSS richness of more than 20 galaxies, 
corresponding roughly to APM type clusters. 
This estimated richness corresponds
to the number of galaxies having $r$-band magnitudes $\le m_{3}+2$ (where $m_{3}$ is 
the magnitude of the third brightest cluster member) 
within a radius of $0.75 \;h^{-1}$ Mpc, 
estimated using the approximate redshift of the cluster.   

Comparing our SMPMFA sample with the 29
clusters of Goto et al. (2002), we find 5 in common.

\subsubsection{Cluster Shapes}
We investigate the reality of our candidate clusters by
fitting a King's profile to their projected galaxy distribution. We expect
that a galaxy enhancement produced by projection effects 
should have a rather flat profile and thus the goodness of fit should
be low\footnote{Note however that a real cluster with significant substructures
could also produce a rather poor King's profile}. 
The King's profile is given by:
\begin{equation}\label{eq:PROF}
\Sigma(\theta) \propto
\left[ 1+\left(\frac{\theta}{\theta_{c}}\right)^{2} \right]
^{-\alpha} \;\;, 
\end{equation}
where $\theta_{c}$ is the angular cluster core radius. 
The slope $\alpha$ is related to $\beta$, 
the ratio of the specific energy in galaxies to 
the specific thermal energy in the gas, by $\alpha=(3\beta-1)/2$ and
spans the range $0.6 \mincir \alpha \mincir 1$
(cf. Bahcall \& Lubin 1993; Girardi et al. 1995; Girardi et al. 1998).

In order to quantify the parameters of the fit 
we perform a standard $\chi^{2}$ 
minimization procedure between the measured  
surface density of each of our clusters and eq.(\ref{eq:PROF}):
\be
\chi^{2}=\sum_{i=1}^{n} \left[ 
\frac{\Sigma_{\rm data}^{i}(\theta)-\Sigma_{\rm King}^{i}
(\theta,\theta_{c},\alpha)}{\sigma^{i}}\right]^{2} \;\;.
\ee 
In Table 1 we present our cluster positions and the parameters of the King's
profile, as well as the $\chi^{2}$ probability ($P_{\chi^{2}}$) of
consistency between the King's profile and the cluster density
profile. The errors are $2\sigma$ estimates, corresponding to
$\Delta \chi^{2}=6.17$ ($\Delta \chi^{2}\equiv \chi^{2}-\chi_{\rm min}^{2}$, 
with $\chi_{\rm min}^{2}$ the absolute minimum value of 
the $\chi^{2}$). 
All but one candidate clusters are 
well fitted by a King's profile with
$P_{\chi^{2}}\ge 0.14$, while the slope $\alpha$ of the profile is
$\simeq 0.7$, corresponding to a $\beta$-parameter of $\simeq 0.8$ 
(cf. Cavaliere \& Fusco-Femiano 1976; Jones \& Forman
1999). Furthermore, we present the estimated cluster redshifts (either
from Goto et al. 2000 or from the MFA method). In the case of cluster
\# 6 we give the spectroscopic redshift from Couch et al. (1991).

For those clusters which are well fitted by a King-like profile
we can give an approximation regarding their spatial core radius $(r_{c})$. 
Different studies (Girardi et al. 1995; Girardi et al. 1998), 
have shown that the radial core radius of Abell clusters has
an average value of $\bar{r_{c}} \simeq 0.10-0.15 \; h^{-1}$ Mpc.
Therefore, having an estimate of the cluster redshift,
the corresponding core radius can be easily found from
$r_{c}=d_{\rm L} \tan(\theta_{c})$, where $d_{\rm L}$ is the luminosity
cluster distance.

In order to check for possible evolutionary
trends regarding the cluster core, we have plotted the
core radius as a function of redshift. As shown in Fig.4 we find
a significant correlation; the coefficient is $\sim 0.60$ with 
the probability of zero correlation being $\le 10^{-5}$.
In the local universe ($0\le z \le 0.15$), the core radius is approximately equal with
$\sim 0.10 h^{-1}$Mpc, which is similar to the value 
derived by Girardi et al. (1998). The line of the best fit is given by
\be
r_{c}\simeq 0.530(\pm0.291)z+0.138(\pm 0.134) \;\;.
\ee  
\begin{figure}
\mbox{\epsfxsize=7.5cm \epsffile{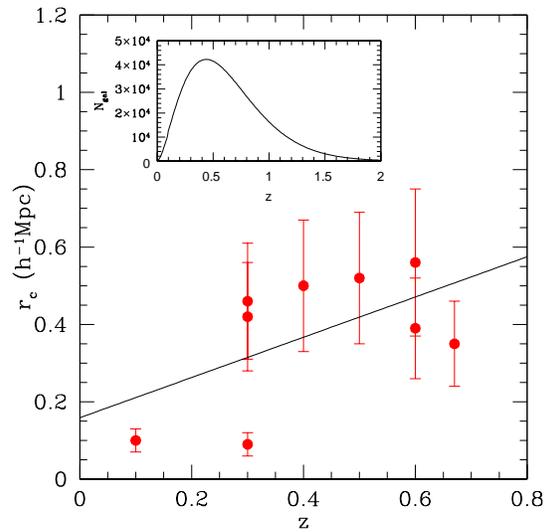}}
\caption{The cluster core radius ($r_{c}$) redshift correlation 
for the analyzed SMPMFA. In the insert we show the SDSS redshift
selection function using the Blanton et al. (2003) luminosity function.}
\end{figure}
In order to check if the observed evolutionary trend is an artifact of
possible systematic effects, we
have firstly investigated the effect of the SDSS redshift selection function.
We have determined the predicted SDSS redshift distribution using the
$r$-band luminosity function of Blanton et al. (2003) 
for galaxies between $0.02\le z \le 0.22$. We have 
extrapolated their luminosity dependent 
density evolution model up to $z=2$ and find
that the SDSS redshift distribution increases 
up to $z\sim 0.45$ (insert of Fig. 4). This implies that the sample 
is roughly volume limited out to this distance and thus 
the $r_c - z$ trend seen in Fig.4
should not be due to systematic effects related to undersampling of
the cluster galaxy population.
Since we have used a single band (r-band) to detect clusters,
we also explore the effect of determining their morphological
characteristics in different rest-frame bands. We note that
the wavelength difference (effective $\lambda$) between the Sloan $r$
and $z$ bands is $\sim$2800 A, which corresponds to a redshift
difference of $\sim$0.45. Therefore, the rest-frame r-band at low redshifts
($z\lesssim 0.15$) roughly shifts to the z-band at $z \sim$ 0.6. We then compare the mean
core-radius of our most distant clusters (at $\sim$0.6 and 0.67) using the
z-band (within its completness limit $m_z \leq 21.5$)
with the mean core-radius of our relatively nearby clusters
($z<0.3$) determined in the r-band (within its the completness limit of
$m_r \leq 22$). We find that the r-band core radius of the 4
low-$z$ clusters is 0.27 $\pm 0.19$ while the z-band core radius of
the 2 high-$z$ clusters is 0.65 $\pm0.35$. Therefore it seems that the effect of the different
rest-frame bands in the definition of the cluster morphology (at least
for our sample and the redshifts covered) does neither create nor mask the observed
evolutionary trend.

We list in Table 1 the cluster shape parameters
determined by using the moments of 
inertia method (cf. Carter \& Metcalfe 1980; 
 Basilakos, Plionis, Maddox, 2001).
This method is based in evaluating the moments
$I_{11}=\sum\ \delta_{j}(r_{j}^{2}-x_{j}^{2})$,
$I_{22}=\sum\ \delta_{j}(r_{j}^{2}-y_{j}^{2})$,
$I_{12}=I_{21}=-\sum\ \delta_{j} x_{j} y_{j}$, where $\delta_{j}$
is the galaxy overdensity of each cell and $x, y$ are the Cartesian
coordinates of each cell, after transforming  
the equatorial coordinates into an equal area
coordinate system, centered on the cluster center.
Then, diagonalizing the inertia tensor
\begin{equation}\label{eq:diag}
{\rm det}(I_{ij}-\lambda^{2}M_{2})=0 \;\;\;\;\; {\rm (M_{2} \;is \; 
2 \times 2 \; unit \; matrix.) }
\end{equation}
we obtain the eigenvalues $\lambda_{1}$, $\lambda_{2}$, from which we
define the ellipticity of the configurations under study by:
$\epsilon=1-\lambda_{2}/\lambda_{1}$, with $\lambda_{1}>\lambda_{2}$.
The corresponding eigenvectors provide the direction of the principal axis.
It is evident that 3 out of 9 clusters have large projected 
ellipticities ($\epsilon > 0.5$).

\section{Comparing Clusters in the Optical \& X-ray bands}
\subsection{X-ray Cluster Detection}
The full details of the cluster detection procedure and the 
cluster X-ray luminosity and temperature determinations
will be described
in Gaga et al. ({\em in preparation}). Here we only briefly 
present the main method and results.

In order to detect candidate clusters in our XMM fields
we use the soft 0.3-2 keV data since this band maximizes the
signal to noise ratio especially in the case of relatively
low temperature galaxy clusters. In particular we 
use the {\sl EWAVELET} detection algorithm of the {\it XMM-Newton}
SAS v.5.3 analysis software package, which 
detects sources on the wavelet transformed images.

We search for sources down to the 5$\sigma$ detection threshold,
on the PN and the co-added MOS1 and MOS2 detector images, separately.
The output list is fed into the SAS {\sl EMLDETECT} algorithm
which performs a maximum likelihood PSF fit on
each source  yielding a likelihood for the extension.
We are using an extension likelihood threshold
corresponding to a level of less than $\sim$ 0.5\% spurious extended sources.
We have detected 7 candidate clusters on the MOS mosaic, while
5 extended sources were detected on the PN images, out of which 3
overlap with the MOS candidates. 

We have also visually inspected the candidate
clusters and found that in some occasions the high extension
likelihood was due to multiple point sources (confusion). 
These spurious detections are excluded
from our final X-ray selected cluster candidate list. This results 
in a final list of 4 X-ray cluster candidates, out of which 2 are
detected on both the
MOS and PN, one is detected only on the MOS (there was no valid PN image)
and the last one is detected only on the PN (note that this cluster
was detected with a slightly smaller likelihood probability, $\sim 0.99$).
The faintest extended source has a flux of 
$\rm \sim 2 \times 10^{-14}erg ~cm^{-2}~s^{-1}$.
Note that for the (0.3-2) keV energy band the
net X-ray count rate for our
detected extented sources is 3 to  $7 \times 10^{-3}$cts s$^{-1}$,
within a 30$^{''}$ radius circle. 

In Table 2 we present the properties of the X-ray cluster
candidates. 
\begin{figure}
\mbox{\epsfxsize=7.5cm \epsffile{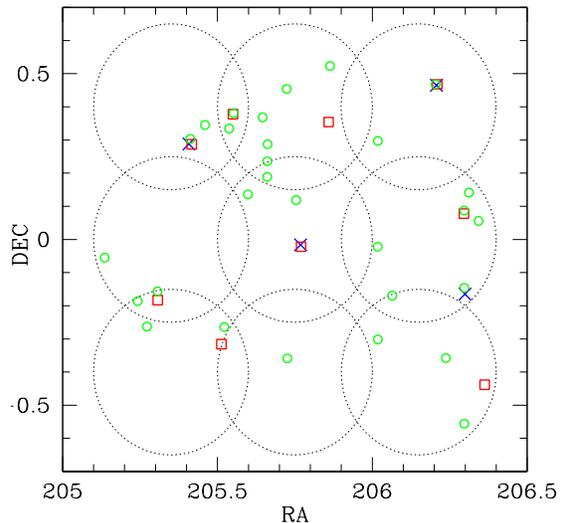}}
\caption{
The candidate cluster positions within our shallow
XMM-{\it Newton} survey. Open squares are the optically selected clusters 
using the SMPMFA technique while open circles are the 
Goto et al. (2002) clusters. Crosses are X-ray selected clusters of
Gaga et al. (in preparation 2004). The large dotted circles represent the 
15 arcmin radius XMM fields of view covered by our shallow XMM survey.}
\end{figure}

\begin{table*}
\caption[]{The X-ray cluster properties: (1) and (2) equitorial
  coordinates (J2000) (2) X-ray Luminosity ($\rm erg~s^{-1}$) in the 0.3-2 keV band, (3) X-ray flux
(0.3-2) keV, in units of 10$^{-14}$erg sec$^{-1}$, 
cm$^{-2}$, (4) cluster temperature from Gaga et al. (in preparation 2004),
(5) the XMM field, (6) optical follow up and finally, 
the cluster redshift. 
Note that the subscripts 1,2 and 4 are the same as in Table 1.}
\tabcolsep 8pt
\begin{tabular}{ccccccccc} \hline
$\alpha$ & $\delta$ & $L_{x}$     &  $f_x$    &  $T (keV)$ &
XMM-Field&Optical&$z$ & detector \\  \hline
 13 41 39.1 & +00 17 39.3  & $10^{42.63}$	 &1.86     & $\sim$3& F864-1&
SMPMFA+CE&$0.40^{2}$ & PN+MOS \\
 13 43 04.8 & -00 00 56.3  & $10^{43.60}$	 &5.55     & $\sim$ 2& F864-5&
SMPMFA&$0.67^{4}$ & PN+MOS \\
 13 45 11.9 & -00 09 52.6  & $10^{42.00}$	 &5.05     &$\sim$ 1& F864-6&
   CE&$0.12^{1}$ & MOS \\
 13 44 49.7 & +00 27 53.3  & $10^{42.87}$	 &6.03     &$\sim$ 1& F864-3&
    SMPMFA+CE&$0.30^{2}$ & PN \\
%
\hline
\end{tabular}				 
\end{table*}

\subsection{The optical versus X-ray cluster detections}
We perform a cross-correlation between the different 
cluster catalogues (optical, X-rays), using a 3 arcmin 
matching radius to identify common objects.
In Fig. 5, we plot the positions
of (a) our optically detected cluster candidates (SMPMFA sample), 
(b) the Goto et al. (2002) clusters 
and (c) the
X-ray detected XMM-{\em Newton} clusters.
All four X-ray detections coincide with optical cluster candidates from different
methods, with the largest coincidence rate (3 out of 9) being with our SMPMFA optical candidates. 
The most distant cluster in our SMPMFA sample, located at $\alpha =$13h
43min 5 sec
$\delta=$ 00$^{o}$ 00$^{'}$ 56$^{''}$, found also in X-rays, is missed by Goto et al. (2002).
\begin{table*}
\caption[]{Cross correlations results. The correspondence 
of the columns is as follows:
comparison pair and the number of common clusters.
}
\tabcolsep 9pt
\begin{tabular}{lc} 
\hline
Methods compared & Number of common clusters\\ \hline 
SMPMFA - CE & 5\\
SMPMFA - X-rays & 3\\
CE - X-rays & 3 \\
\hline
\end{tabular}
\end{table*}

The cross-correlation results between the different cluster detection
methods are presented in Table 3, which in the
first column lists the comparison method pair 
and in the second column the number of common objects.

It is interesting that our SMPMFA combined method finds 3 out of the 4
X-ray clusters, missing the nearest one, which from its X-ray
luminosity appears to be a group of galaxies, rather than a cluster.
However, there are still 6 SMPMFA clusters that do not appear to have
X-ray counterparts, for which we give their $3\sigma$ flux
upper limits (see Table 1 last column). Conversion from
count rate to flux was done assuming a bremsstrahlumg spectrum with
3 keV. The flux was measured in a $30^{''}$ radius cell and 
the correction to the total flux was performed taking into acount
a King profile with $r_c = 150 \;h^{-1}$ kpc.
This could be a hint that 
these clusters are either the results of projection effects 
(which however cannot explain
the good King profile fits - see Table 1), or that our XMM-{\em Newton}
survey is too shallow to reveal the probably weak X-ray emission from these
clusters.

In order to address this issue and to study the relation 
between the limiting flux of our 
X--ray survey with respect to exposure time, we have carried out the following 
experiment. We have analysed observations taken from 15 XMM-{\em Newton} public fields with 
mean exposure times $\sim 21$ ksec in the soft 0.3-2 keV
band, after filtering to correct for high particle background periods.  
Using the parameters of the SAS software as described previously, we have 
detected 31 candidate clusters. The faintest cluster detected, with a
flux of $\sim 5 \times 10^{-15}$
erg cm$^{-2}$ s$^{-1}$, was found in the deepest field with an exposure
time of 37 ksec. We then reduce the exposure times 
to a new mean value of $\sim 5$ ksec, similar to our shallow survey,
and find only 9 out of the 31 previously identified candidate 
clusters (29\%), having a limiting flux of $\sim 2.3 \times 10^{-14}$ erg cm$^{-2}$ 
s$^{-1}$. Therefore, had we had deeper XMM-{\em Newton} observations (by
an average factor of $\sim 5$ in exposure time) we would 
have detected $\sim 13$ X-ray candidate
clusters in the region covered by our shallow XMM-{\em Newton}/2dF survey, which is 
consistent (within $1\sigma$) with the number of our SMPMFA cluster
candidates. 

\section{Conclusions}
We have made a direct comparison between optical and X-ray based techniques
used to identify clusters. 
We have searched for extended emission in our shallow XMM-{\it Newton} 
Survey, which covers a $\sim 1.6 \;\;{\rm deg^{2}}$ area (8 out of 9
original XMM pointings) 
in the North Galactic Pole region and we have 
detected 4 candidate X-ray clusters.
We have then applied a new cluster finding algorithm on the SDSS galaxy
distribution in this region, which is based on merging two independent 
selection methods - a smoothing-percolation 
SMP technique, and a Matched Filter 
Algorithm (MFA). Our final optical cluster catalogue, called the SMPMFA list, 
counts 9 candidate clusters with richness of more than 20 galaxies, 
corresponding roughly to the APM richness limit. 

Out of the 4 X-ray candidate clusters 3 are common with our SMPMFA
list. 
This relatively, small number of optical SMPMFA 
cluster candidates observed in X-rays
suggest that some of the optical cluster 
candidates are either projection effects or
poor X-ray emitters  and hence they are fainter in X-rays 
than the limit of our shallow survey 
$f_{x}(0.3-2 keV) \rm \simeq 2 \times 10^{-14}erg ~cm^{-2}~s^{-1}$.
This latter explanation seems to be supported from an analysis of 
public XMM-{\em Newton} fields with larger exposure times.

\section*{Acknowledgments}
We thank an anonymous referee for many useful comments that help
improve this work.
This research was jointly funded by the European Union
and the Greek Government in the framework 
of the project {\em 'X-ray Astrophysics with ESA's mission XMM'}.
within the program 'Promotion
of Excellence in Technological Development and Research'. 
MP also acknowledges funding by the Mexican Government grant
No. CONACyT-2002-C01-39679.


{\small
\begin{thebibliography}{}
\bibitem[]{}Abell, G.O., Corwin, H.G., Olowin, R.P., 1989, ApJS, 70, 1
\bibitem[]{}Bahcall, N.A., 1988, ARA\&A, 26, 631
\bibitem[]{}Bahcall, N.A., \& Lubin, L. M., 1993, ApJ, 415, L17
\bibitem[]{}Bahcall, N.A., et al., 2003, ApJS, in press, astro-ph/0305202
\bibitem[]{}Basilakos S., Plionis M., Maddox S. J., 2001, MNRAS, 316, 779
\bibitem[]{}Bertin, E., \&, Arnouts, S., 1996, A\&AS, 117, 393
\bibitem[]{}Blanton M.R., et al., 2003, ApJ, 592, 819
\bibitem[]{}B\"{o}hringer H., 1995, Large Scale Structure in the Universe, Proc. of an International Workshop, 
Potsdam Germany, 18-24 September 1994. Ed. by 
Jan P. M\"{u}cket, Stefan Gottloeber, 
and Volker M\"{u}ller. Singapore: World Scientific, 1995, p.181
\bibitem[]{}B\"{o}hringer H., et al. , 2001, A\&A, 369, 826
\bibitem[]{}Borgani, S., \&, Guzzo, L., 2001, Nature, 409, 39
\bibitem[]{}Bruzual A.G., \&, Charlot, S. 1993, ApJ, 405, 538
\bibitem[]{}Carlberg, R.G., Yee, H.K.C., Ellingson, E.,, Abraham, R., Gravel, P., Morris, S.,
Pritchet, C. J., 1996, ApJ, 369, 16
\bibitem[]{}Carter, D. \& Metcalfe, J., 1980, MNRAS, 191, 325
\bibitem[]{}Castander, F.J., et al., 1995, Nature, 377, 39
\bibitem[]{}Cavaliere, A.,\&, Fusco-Femiano R., 1976, A\&A, 49, 137
\bibitem[]{}Couch, W.J., Ellis, R. S., MacLaren, I., Malin, D. F., 1991, MNRAS, 249, 606
\bibitem[]{}Dalton, G.B., Efstathiou, G., Maddox, S. J., Sutherland, W. J., 1994, MNRAS, 269, 151
\bibitem[]{}Donahue, M., et al., 2002, ApJ, 569, 689
\bibitem[]{}Ebeling, H., et al., 1996a, Proc. Roentgenstrahlung from the Universe, MPE, Report 263,
ed. H. U. Zimmerman, J. Tr\"{u}mper, H. Yorke (ISSN 0178-1719), 579
\bibitem[]{}Ebeling, H., Voges, W., B\"{o}hringer H., Edge, A. C., Huchra, J. P., Briel, U. G.,
1996B, MNRAS, 281, 799
\bibitem[]{}Ebeling, H., et al., 2000, ApJ, 534, 133
\bibitem[]{}Frenk, C. S., White, S. D. M., Efstathiou, G., Davis, M., 1990, ApJ, 351, 10
\bibitem[]{} Fukugita, M., Shimasaku, K., Ichikawa, T. 1995, PASP,
107, 945
\bibitem[]{}Georgakakis, A., et al., 2004, MNRAS, in press ({\em astro-ph/0305278})
\bibitem[]{} Gioia, I. M., Henry, J. P., Mullis, C. R., Voges, M., Briel,
U. G., B\"{o}hringer H., Huchra, J. P., 2001, ApJ, 553, L105
\bibitem[]{}Girardi, M., Biviano A., Giuricin G., Mardirossian F.,
Mezzetti M., 1995, ApJ, 438, 527
\bibitem[]{}Girardi, M., Giuricin G., Mardirossian, F., Mezzetti M.,
Boschin W., 1998, ApJ, 505, 74
\bibitem[]{}Gladders, M. D., \&, Yee, H. K. C., 2000, AJ, 120, 2148
\bibitem[]{}Goto, T., et al., 2002, AJ, 123, 1807
\bibitem[]{} Jones, C., Forman, W., 1999, ApJ, 511, 65 
\bibitem[]{}Lidman, C. E., \&, Peterson, B. A., 1996, AJ, 112, 2454
\bibitem[]{}Lumsden, S. L., Nichol, R.C ., Collins, C. A., Guzzo, L., 1992, MNRAS, 258, 1
\bibitem[]{}Nichol, R. C., 2002, in ASP Conf. Ser., Tracing Cosmic Evolution with
Galaxy Clusters, ed. S. Borgani, M. Mezzeti, R. Valdarnini,  268, 57
\bibitem[]{}Olsen, L. F., et al., 1999, A\&A, 345, 681
\bibitem[]{}Postman, M., Lubin, L.M., Gunn, J. E., Oke, J.B.,
  Hoessel, J.G., Schneider, D.P.,Christensen, J.A., 1996, AJ, 111, 615
\bibitem[]{}Pozzetti, L., Bruzual A., G., Zamorani, G., 1996, MNRAS, 281, 953
\bibitem[]{}Rosati, P., Borgani, S., Norman, C., 2002, ARA\&A, 40, 539
\bibitem[]{}
Scharf, C. A., Jones, L. R., Ebeling, H., Perlman, E., Malkan, M., Wegner, G., 1997, ApJ, 477, 79
\bibitem[]{}
Stocke, J. T., Morris, S. L., Gioia, I. M., 
Maccacaro, T., Schild, R., Wolter, A., Fleming, T. A., Henry, J. P., 1991, 
ApJS,76,813
\bibitem[]{}Stoughton, C., et al., 2002, AJ, 123, 485
\bibitem[]{} Sutherland, W., 1988, MNRAS, 234, 159
\bibitem[]{}West, M. J., Jones, C., Forman, W., 1995, ApJ, 451, L5
\bibitem[]{}York, D. G., et al., 2000, AJ, 120, 1579
\bibitem[]{}Zwicky, F., Herzog, E., Wild, P., 1968, Pasadena: California Institute of Technology (CIT),
1961-1968



\end{thebibliography}
}
\end{document}